\documentclass[showpacs,reprint]{revtex4}
\usepackage{amsfonts}
\usepackage{amssymb}
\usepackage{amsmath}
\usepackage{color}
\usepackage{hyperref}
\usepackage{latexsym}
\usepackage[dvips]{graphicx}

\usepackage{epsf}
\usepackage{multirow}

\begin{document}

\title{Dark energy from the gas of wormholes}
\author{ A. A. Kirillov}
\email{ka98@mail.ru}
\author{E.P. Savelova}
\email{sep$\_$ 22.12.79@inbox.ru} \affiliation{Dubna International
University of Nature, Society and Man, Universitetskaya Str. 19,
Dubna, 141980, Russia }
\date{}
\pacs{
}

\begin{abstract}
We assume the space-time foam picture in which the vacuum is filled with a
gas of virtual wormholes. It is shown that virtual wormholes form a finite
(of the Planckian order) value of the energy density of zero-point
fluctuations. However such a huge value is compensated by the contribution
of virtual wormholes to the mean curvature and the observed value of the
cosmological constant is close to zero. A non-vanishing value appears due to
the polarization of vacuum in external classical fields. In the early
Universe some virtual wormholes may form actual ones. We show that in the
case of actual wormholes vacuum polarization effects are negligible while
their contribution to the mean curvature is apt to form the observed dark
energy phenomenon. Using the contribution of wormholes to dark matter and
dark energy we find estimates for characteristic parameters of the gas of
wormholes.
\end{abstract}

\maketitle

\section{Introduction}

As is well known modern astrophysics (and, even more generally, theoretical
physics) faces two key problems. Those are the nature of dark matter and
dark energy. Recall that more than $90\%$ of matter of the Universe has a
non-baryonic dark (to say, mysterious) form, while lab experiments still
show no evidence for the existence of such matter. Both dark components are
intrinsically incorporated in the most successful $\Lambda $CDM (Lambda cold
dark matter) model which reproduces correctly properties of the Universe at
very large scales (e.g., see \cite{Pr} and references therein). We point out
that $\Lambda $CDM predicts also the presence of cusps ($\rho _{DM}\sim 1/r$%
) in centers of galaxies \cite{Cusp} and a too large number of galaxy
satellites. Therefore other models are proposed, e.g., like axions \cite{M09}%
, which may avoid these. To be successful such models should involve a
periodic self-interaction and therefore require a fine tuning, while in
general the presence of standard non-baryonic particles cannot solve the
problem of cusps. Indeed, if we admit the existence of a self-interaction in
the dark matter component, or some coupling to baryonic matter (which should
be sufficiently strong to remove cusps), then we completely change
properties of the dark matter component at the moment of recombination and
destroy all successful predictions at very large scales. Recall that both
warm and self-interacting dark matter candidates are rejected by the
observing $\Delta T/T$ spectrum \cite{Pr}. In other words, the two key
observational phenomena (cores of dark matter in centers of galaxies \cite%
{Core} and $\Delta T/T$ spectrum) give a very narrow gap for dark matter
particles which seems to require attracting some exotic objects in addition
to standard non-baryonic particles.

As it was demonstrated recently \cite{KS10} the problem of cusps can be
cured, if some part of non-baryon particles is replaced by wormholes.
Wormholes represent extremely heavy (in comparison to particles) objects
which at very large scales behave exactly like non-baryon cold particles,
while at smaller scales (in galaxies) they strongly interact with baryons
and form the observed \cite{Core} cored ($\rho _{DM}\sim const$)
distribution. We note that stable wormholes violate necessarily the averaged
null energy conditions which gives the basic argument against of the
existence of such objects. Without exotic matter stable wormholes may
however exist in modified theories, e.g., see \cite{KKK12} and references
therein. In the case when the energy conditions hold a wormhole collapses
into a couple of conjugated (of equal masses) blackholes which almost
impossible to distinguish from standard primordial blackholes. However, the
topological non-triviality of such objects retains and gravitational effects
of a gas of wormholes considered in \cite{KS07} and some results of \cite%
{KS10} still remain valid which means that non-traversable wormholes can be
used to smooth cusps in centers of galaxies. Thus, it worth expecting that
wormholes may play an important role in the explanation of the dark matter
phenomenon.

Save the dark matter component $\Lambda $CDM requires the presence ( $\sim
70\%$) of dark energy (of the cosmological constant). Moreover, there is
evidence for the start of an acceleration phase in the evolution of the
Universe \cite{Ac}. In the present paper we use virtual wormholes to
estimate the contribution of zero-point fluctuations in the value of the
cosmological constant. The idea to relate virtual wormholes (or baby
universes) and the cosmological constant is not new, in somewhat different
context it was used by Coleman in \cite{col} and developed in \cite{bab}.
Our basic aim is to demonstrate that virtual wormholes form a finite value
of the energy density of zero-point fluctuations.

It is necessary to point here out to the principle difference between actual
and virtual wormholes. The principle difference is that a virtual wormhole
exists only for a very small period of time and at very small scales, and
does not necessary obey to the Einstein equations. Therefore, the averaged
null energy condition (ANEC) cannot forbid the origin of such an object. For
the future we also note that a set of virtual wormholes may work as an
actual wormhole opening thus the way for an artificial construction of
wormhole-type objects in lab experiments.

In the present paper we describe a virtual wormhole as follows. From the
very beginning we use the Euclidean approach (e.g., see \cite{H78} and the
standard textbooks \cite{book}). Then the simplest virtual wormhole is
described by the metric ($\alpha =1,2,3,4$)%
\begin{equation}
ds^{2}=h^{2}\left( r\right) \delta _{\alpha \beta }dx^{\alpha }dx^{\beta },
\label{wmetr}
\end{equation}%
where
\begin{equation}
h\left( r\right) =1+\theta \left( a-r\right) \left( \frac{a^{2}}{r^{2}}%
-1\right)
\end{equation}%
and $\theta \left( x\right) $ is the step function. Such a wormhole has
vanishing throat length, while the step function at the junction may cause a
problem in Einstein's equation or when a topological Euler term is involved.
A more careful analysis needs to consider distributional curvature etc., see
\cite{taub}. To avoid these difficulties we may consider from the very
beginning a wormhole of a finite throat length $\sim 1/\beta $ where the
step function is replaced with a smooth function (e.g., $\theta \left(
x,\beta \right) =(\exp (\beta x)+1)^{-1}$ ). Then where it is necessary one
may consider the limit $\beta \rightarrow \infty $ only in final
expressions. This insures that the Bianchi identity holds and that the above
metric remains inside of the domain of usual gravity.

In the region $r>a$, $h=1$ and the metric is flat, while the region $r<a$,
with the obvious transformation $y^{\alpha }=\frac{a^{2}}{r^{2}}x^{\alpha }$%
, is also flat for $y>a$. Therefore, the regions $r>a$ and $r<a$ represent
two Euclidean spaces glued at the surface of a sphere $S^{3}$ with the
centre at the origin $r=0$ and radius $r=a$. Such a space can be described
with the ordinary double-valued flat metric in the region $r_{\pm }>a$ by
\begin{equation}
ds^{2}=\delta _{\alpha \beta }dx_{\pm }^{\alpha }dx_{\pm }^{\beta },
\label{wmetr2}
\end{equation}%
where the coordinates $x_{\pm }^{\alpha }$ describe two different sheets of
space. Now, identifying the inner and outer regions of the sphere $S^{3}$
allows the construction of a wormhole which connects regions in the same
space (instead of two independent spaces). This is achieved by gluing the
two spaces in (\ref{wmetr2}) by motions of the Euclidean space (the Poincare
motions). If $R_{\pm }$ is the position of the sphere in coordinates $x_{\pm
}^{\mu }$, then the gluing is the rule%
\begin{equation}
x_{+}^{\mu }=R_{+}^{\mu }+\Lambda _{\nu }^{\mu }\left( x_{-}^{\nu
}-R_{-}^{\nu }\right) ,  \label{gl}
\end{equation}%
where $\Lambda _{\nu }^{\mu }\in O(4)$, which represents the composition of
a translation and a rotation of the Euclidean space (Lorentz
transformation). In terms of common coordinates such a wormhole represents
the standard flat space in which the two spheres $S_{\pm }^{3}$ (with
centers at positions $R_{\pm }$) are glued by the rule (\ref{gl}). We point
out that the physical region is the outer region of the two spheres. Thus,
in general, the wormhole is described by a set of parameters: the throat
radius $a$, positions of throats $R_{\pm }$, and rotation matrix $\Lambda
_{\nu }^{\mu }\in O(4)$.

In the present paper we assume the space-time foam picture in which the
vacuum is filled with a gas of virtual wormholes. We show that virtual
wormholes form a finite (of the Planckian order) value of the energy density
of zero-point fluctuations. However such a huge value is compensated by the
contribution of virtual wormholes to the mean curvature and the observed
value of the cosmological constant should be close to zero.

To achieve our aim we, in section II, present the construction of the
generating functional in quantum field theory. The main idea is that the
partition function includes the sum over field configurations and the sum
over topologies. Where the sum over topologies is the sum over virtual
wormholes described above. Such an approach gives a rather good leading
approximation for calculation of the partition function and corresponds to
the standard methods (e.g., Ritza method, etc.). In section III we
investigate properties of the two-point Green function. We show that the
presence of the gas of virtual wormholes can be described by the topological
bias exactly as it happens in the presence of actual wormholes \cite%
{KS07,KS10}. For limiting topologies when the density of virtual wormholes
becomes infinite the Green function shows a good ultraviolet behavior which
means that there exists a class of such systems when quantum field theories
are free of divergencies. We demonstrate how the sum over topologies defines
the mean value for the bias which takes the sense of a cutoff function in
the space of modes. In section IV we explicitly demonstrate that for a
particular set of virtual wormholes the bias defines not more than the
projection operator on the subspace of functions obeying to the proper
boundary conditions at wormhole throats. The projective nature of the bias
means that wormholes merely cut some portion of degrees of freedom (modes).
Phenomenologically it means that wormholes can be described by the presence
of ghost fields which compensate the extra (cut by wormholes) modes. In
section V we show how the cutoff expresses via some dynamic parameters of
wormholes. The exact definition of such parameters we leave for the future
investigation. In section VI we consider the origin of the cosmological
constant. We demonstrate that the cosmological constant is determined by the
contribution of the energy density of zero-point fluctuations and by the
contribution of virtual wormholes to the mean curvature. We estimate
contribution of virtual wormholes to the mean curvature and show also that
wormholes lead to a finite (of the planckian order) value of $\left\langle
T_{\mu \nu }\right\rangle $ which requires considering the contribution from
the smaller and smaller wormholes with divergent density $n\rightarrow
\infty $. We also present arguments of why in the absence of external
classical fields the total value of the cosmological constant is exactly
zero, while it acquires a non-vanishing value due to vacuum polarization
effects (i.e., due to an additional distribution of virtual wormholes) in
external fields. We also speculate on the possibility of the formation of
actual wormholes and in section VII estimate their contribution to the dark
energy. Finally in section VIII we repeat basic results an discuss some
perspectives.

\section{Generating function}

The basic aim of this section is to construct the generating functional
which can be used to get all possible correlation functions. Consider the
partition function which includes the sum over topologies and the sum over
field configurations
\begin{equation}
Z_{total}=\sum\limits_{\tau }\sum\limits_{\varphi }e^{-S}.  \label{z1}
\end{equation}%
For the sake of simplicity we use from the very beginning the Euclidean
approach. The action has the form
\begin{equation}
S=-\frac{1}{2}\left( \varphi \hat{A}\varphi \right) +\left( J\varphi \right)
\label{act}
\end{equation}%
and we use the notions $\left( J\varphi \right) =\int J\left( x\right)
\varphi \left( x\right) d^{4}x$. If we fix the topology of space by placing
a set of wormholes with parameters $\xi _{i}$, then the sum over field
configurations $\varphi $ gives the well-known result
\begin{equation}
Z^{\ast }\left( J\right) =Z_{0}(\hat{A})e^{-\frac{1}{2}\left( J\hat{A}%
^{-1}J\right) },  \label{gf2}
\end{equation}%
where $Z_{0}(\hat{A})=\int \left[ D\varphi \right] e^{\frac{1}{2}\left(
\varphi \hat{A}\varphi \right) }$ is the standard expression and $\hat{A}%
^{-1}=A^{-1}( \xi) $ is the Green function for a fixed topology, i.e., for a
fixed set of wormholes $\xi _{1},...,\xi _{N}$ .

Consider now the sum over topologies $\tau $. To this end we restrict with
the sum over the number of wormholes and integrals over parameters of
wormholes:
\begin{equation}
\sum\limits_{\tau }\rightarrow \sum\limits_{N}\int
\prod\limits_{i=1}^{N}d\xi _{i}=\int \left[ DF\right] ,  \label{ts}
\end{equation}%
where
\begin{equation}
F\left( \xi ,N\right) =\frac{1}{N}\sum\limits_{i=1}^{N}\delta \left( \xi
-\xi _{i}\right)  \label{F}
\end{equation}%
and $NF$ is the density of wormholes in the configuration space $\xi $. We
also point out that in general the integration over parameters is not free
(e.g., it obeys the obvious restriction $\left\vert \vec{R}_{i}^{+}-\vec{R}%
_{i}^{-}\right\vert \geq 2a_{i}$). This defines the generating function as
\begin{equation}
Z_{total}\left( J\right) =\int \left[ DF\right] Z_{0}(\hat{A})e^{-\frac{1}{2}%
\left( J\hat{A}^{-1}J\right) }.  \label{ts2}
\end{equation}%
The sum over topologies assumes an additional averaging out for all mean
values with the measure $d\mu _{N}=\rho \left( \xi ,N\right) d^{N}\xi $,
where
\begin{equation}
\rho \left( \xi ,N\right) =\frac{Z_{0}(\hat{A}\left( \xi ,N\right) )}{%
Z_{total}\left( 0\right) }  \label{meas}
\end{equation}%
which obey the obvious normalization condition $\sum_{N}\int d\mu
_{N}=\sum_{N}\rho _{N}=1$. The averaging out over topologies assumes the two
stages. First we fix the total number of wormholes~$N$ and average over the
parameters of wormholes $\xi $ (i.e., over parameters of a static gas of
wormholes in $R^{4}$). Then we sum over the number of wormholes $N$ (the
so-called big canonical ensemble).

The basic difficulty of the standard field theory is that the perturbation
scheme based on (\ref{gf2}) leads to divergent expressions. This remains
true for every particular topology of space (i.e., for\ any particular
finite set of wormholes), since there always exists a scale below which the
space looks like the ordinary Euclidean space. What we expect is that the
sum over all possible topologies will remove such a difficulty.

And indeed, the above measure (\ref{meas}) has the structure
\begin{equation*}
Z_{0}(\hat{A} \left( \xi ,N\right) )=\exp \left( -\int \Lambda \left( \xi
,N\right) d^{4}x\right)
\end{equation*}
where $\Lambda \left( \xi ,N\right) $ is the cosmological constant related
to the energy density of zero-point fluctuations calculated for a particular
distribution of wormholes \footnote{%
We recall that the total cosmological constant should include also the
contribution from the mean curvature (\ref{mcv}).}. Any finite distribution
of wormholes leads to the divergent expression $\Lambda \left( \xi ,N\right)
\rightarrow \infty $ and is suppressed (i.e., $\rho \left( \xi ,N\right)
\rightarrow 0$). However, the sum over all possible topologies assumes also
the limiting topologies $n\rightarrow \infty $, where $n=N/V$ is the density
of wormholes. In this limit wormhole throats degenerate into points and the
minimal scale below which the space looks like the Euclidean space is merely
absent. We point out that from the rigorous mathematical standpoint such
limiting topologies cannot be described in terms of smooth manifolds, since
they are not locally Euclidean and does not possess a finite set of maps. In
mathematics similar objects are well known, e.g., fractal sets. However, if
a fractal set is obtained by cutting (by means of a specific rule or
iterations) portions of space, our limiting topologies are obtained by
gluing (identifying) some portions (or in the limit couples of points) of
the Euclidean space. The basic feature of such topologies is that QFT
becomes finite on such a set. Indeed, as we shall see a particular infinite
distribution of wormholes can always be chosen in such a way that the energy
of zero-point fluctuations becomes a finite $0\leq \Lambda _{\infty }\left(
\xi \right) <\infty $ (e.g., see the next section or the second term in (\ref%
{le})). In the sum over topologies only such limiting topologies do survive
(i.e., $\rho _{\infty }\left( \xi \right) \neq 0$).

\section{The two-point Green function}

From (\ref{gf2}) we see that the very basic role in QFT plays the two point
Green function. Such a Green function can be found from the equation
\begin{equation}
\hat{A}G(x,x^{\prime })=-\delta (x-x^{\prime })
\end{equation}%
with proper boundary conditions at wormholes, which gives $G=A^{-1}$. Now
let us introduce the bias function $N(x,x^{\prime })$ as
\begin{equation}
G(x,y)=\int G_{0}(x,x^{\prime })N(x^{\prime },y)dx^{\prime },  \label{gf}
\end{equation}%
where $G_{0}(x,x^{\prime })$ is the ballistic (or the standard Euclidean
Green function) and the bias can be presented as
\begin{equation}
N(x,x^{\prime })=\delta (x-x^{\prime })+\sum\limits_{i}b_{i}\delta (x-x_{i})
\label{m}
\end{equation}%
where $b_{i}$ are fictitious sources at positions $x_{i}$ which should be
added to obey the proper boundary conditions. We point out that the bias can
be explicitly expressed via parameters of wormholes, i.e., $N(x,x^{\prime
})=N(x,x^{\prime },\xi _{1},...\xi _{N})$. For the sake of illustration we
consider first a particular example.

\subsection{The bias for a particular distribution of wormholes (rarefied
gas approximation)}

Consider now the bias for a particular set of wormholes. For the sake of
simplicity we consider the case when $m=0$. The Green function obeys the
Laplace equation
\begin{equation*}
-\Delta G\left( x,x^{\prime }\right) =\delta \left( x-x^{\prime }\right)
\end{equation*}%
with proper boundary conditions at throats (we require $G$ and $\partial
G/\partial n$ to be continual at throats). The Green function for the
Euclidean space is merely $G_{0}\left( x,x^{\prime }\right) =\frac{1}{4\pi
^{2}\left( x-x^{\prime }\right) ^{2}}$ (and $G_{0}\left( k\right) =1/k^{2}$
for the Fourier transform). In the presence of a single wormhole which
connects two Euclidean spaces this equation admits the exact solution. For
outer region of the throat $S^{3}$ the source $\delta \left( x-x^{\prime
}\right) $ generates a set of multipoles placed in the center of sphere
which gives the corrections to the Green function $G_{0}$ in the form (we
suppose the center of the sphere at the origin)%
\begin{equation}
\delta G=-\frac{1}{4\pi ^{2}x^{2}}\sum_{n=1}^{\infty }\frac{1}{n+1}\left(
\frac{a}{x^{\prime }}\right) ^{2n}\left( \frac{x^{\prime }}{x}\right)
^{n-1}Q_{n},
\end{equation}%
where $Q_{n}=\frac{4\pi ^{2}}{2n}\sum_{l=0}^{n-1}\sum_{m=-l}^{l}Q_{nlm}^{%
\ast \prime }Q_{nlm}$ and $Q_{nlm}\left( \Omega \right) $ are
four-dimensional spherical harmonics e.g., see \cite{fock}. In the present
section we shall consider a dilute gas approximation and, therefore, it is
sufficient to retain the lowest (monopole) term only. A single wormhole
which connects two regions in the same space is a couple of conjugated
spheres $S_{\pm }^{3}$ of the radius $a$ with a distance $\vec{X}=\vec{R}%
_{+}-\vec{R}_{-}$ between centers of spheres. So the parameters of the
wormhole are \footnote{%
The additional parameter (rotation matrix $U$) is important only for
multipoles of higher orders.} $\xi =(a,R_{+},R_{-})$. The interior of the
spheres is removed and surfaces are glued together. Then the proper boundary
conditions (the actual topology) can be accounted for by adding the bias of
the source
\begin{equation}
\delta (x-x^{\prime })\rightarrow \delta (x-x^{\prime })~+b\left(
x,x^{\prime }\right) .  \label{b}
\end{equation}%
In the approximation $a/X\ll 1$ (e.g., see for details
\cite{KS07}) the bias for a single wormhole takes the form
\begin{equation}\label{b1}
b_{1}( x,x^{\prime },\xi ) =\frac{a^{2}}{2}\left(
\frac{1}{\left(
R_{-}-x^{\prime }\right) ^{2}}-\frac{1}{\left( R_{+}-x^{\prime }\right) ^{2}}%
\right) \left[ \delta (\vec{x}-\vec{R}_{+})-\delta
(\vec{x}-\vec{R}_{-}) \right] .
\end{equation}%
This form for the bias is convenient when constructing the true Green
function and considering the long-wave limit, however it is not acceptable
in considering the short-wave behavior and vacuum polarization effects.
Indeed, the positions of additional sources are in the physically
non-admissible region of space (the interior of spheres $S_{\pm }^{3}$). To
account for the finite value of the throat size we should replace in (\ref%
{b1}) the point-like source with the surface density (induced on the throat)
i.e.,
\begin{equation}
\delta (\vec{x}-\vec{R}_{\pm })\rightarrow \frac{1}{2\pi ^{2}a^{3}}\delta
(\left\vert \vec{x}-\vec{R}_{\pm }\right\vert -a).  \label{d}
\end{equation}%
Such a replacement does not change the value of the true Green function,
however now all extra sources are in the physically admissible region of
space.

In the rarefied gas approximation the total bias is additive, i.e.,
\begin{equation}
b_{total}\left( x,x^{\prime }\right) =\sum b_{1}\left( x,x^{\prime },\xi
_{i}\right) =N\int b_{1}(x,x^{\prime },\xi )F(\xi )d\xi ,  \label{bs}
\end{equation}%
where $NF$ is given by (\ref{F}). For a homogeneous and isotropic
distribution $F(\xi )=F(a,X)$, then for the bias we find
\begin{equation}
b_{total}\left( x-x^{\prime }\right) =\int \frac{1}{2\pi ^{2}a}\left( \frac{1%
}{R_{-}^{2}}-\frac{1}{R_{+}^{2}}\right)
\delta (\left\vert \vec{x}-\vec{x}%
^{\prime }-\vec{R}_{+}\right\vert -a)NF(\xi )d\xi  \label{bt}
\end{equation}%
Consider the Fourier transform $F\left( a,X\right) =\int F\left(
a,k\right)
e^{-ikX}\frac{d^{4}k}{\left( 2\pi \right) ^{4}}$ and using the integral $%
\frac{1}{x^{2}}=\int \frac{4\pi ^{2}}{k^{2}}e^{-ikx}\frac{d^{4}k}{\left(
2\pi \right) ^{4}}$ we find for $b\left( k\right) =\int b\left( x\right)
e^{ikx}d^{4}x$ the expression
\begin{equation}
b_{total}\left( k\right) =N\int a^{2}\frac{4\pi ^{2}}{k^{2}}\left( F\left(
a,k\right) -F\left( a,0\right) \right) \frac{J_{1}\left( ka\right) }{ka/2}da.
\label{bk}
\end{equation}%
\textit{1. Example of a finite density of wormholes }

Consider now a particular (of a finite density) distribution of wormholes $%
F\left( a,X\right) $, e.g.,
\begin{equation}
NF\left( a,X\right) =\frac{n}{2\pi ^{2}r_{0}^{3}}\delta \left(
a-a_{0}\right) \delta \left( X-r_{0}\right) ,  \label{ff}
\end{equation}%
where $n=N/V$ is the density of wormholes. In the case $N=1$ this function
corresponds to a single wormhole with the throat size $a_{0}$ and the
distance between throats $r_{0}=\left\vert R_{+}-R_{-}\right\vert $. We
recall that the action (\ref{act}) remains invariant under translations and
rotations which straightforwardly leads to the above function. Then $%
NF\left( a,k\right) =\int NF\left( a,X\right) e^{ikx}d^{4}x$ reduces to $%
NF\left( a,k\right) =n\frac{J_{1}\left( kr_{0}\right) }{kr_{0}/2}\delta
\left( a-a_{0}\right) $. Thus from (\ref{bk}) we find
\begin{equation}
b\left( k\right) =-na^{2}\frac{4\pi ^{2}}{k^{2}}\left( 1-\frac{J_{1}\left(
kr_{0}\right) }{kr_{0}/2}\right) \frac{J_{1}\left( ka_0\right) }{ka_0/2}.
\label{b(k)}
\end{equation}%
And for the true Green function we get
\begin{equation*}
G_{true}=G_{0}\left( k\right) N\left( k\right) =G_{0}\left( k\right)
(1+b\left( k\right) ).
\end{equation*}

In the short-wave limit ($ka$, $kr_{0}\gg 1$) $b\left( k\right) \rightarrow
0 $ and therefore $N(k)\rightarrow 1$. This means that at very small scales
the space filled with a finite density of wormholes looks like the ordinary
Euclidean space. In the long-wave limit $k\rightarrow 0$ we get $J_{1}\left(
kr_{0}\right) /\frac{kr_{0}}{2}\approx 1-\frac{1}{2}\left( \frac{kr_{0}}{2}%
\right) ^{2}+...$ which gives $b\left( k\right) \approx -\pi
^{2}na^{2}r_{0}^{2}/2$, while in a more general case we find $b\left(
k\right) \approx -\int \frac{\pi ^{2}}{2}a^{2}r_{0}^{2}n\left(
a,r_{0}\right) dadr_{0}$, where $n\left( a,r_{0}\right) $ is the density of
wormholes with a particular values of $a$ and $r_{0}$, and for the bias
function (\ref{m}) we get
\begin{equation}
N(k)\rightarrow 1-\frac{\pi ^{2}}{2}\int a^{4}n\left( a,r_{0}\right) \frac{%
r_{0}^{2}}{a^{2}}dadr_{0}\leq 1.  \label{mnl}
\end{equation}%
In other words, in the long-wave limit ($ka$, $kr_{0}\ll 1$) the presence of
a particular set of virtual wormholes diminishes merely the value of the
charge values.

\textit{2. Limiting topologies or infinite densities of wormholes}

Consider now the limiting distribution when the density of wormholes $%
n\rightarrow \infty $. Since every throat cuts the finite portion of the
volume $\frac{\pi ^{2}}{2}a^{4}$, this case requires considering the limit $%
a\rightarrow 0$. We assume that in this limit $a^{2}NF(a,X)\rightarrow
\delta \left( a\right) \nu \left( X\right) $ where $\nu \left( X\right) $ is
a finite specific distribution. Then (\ref{bk}) reduces to $b_{total}\left(
k\right) =\frac{4\pi ^{2}}{k^{2}}\left( \widetilde{\nu }\left( k\right) -%
\widetilde{\nu }\left( 0\right) \right) $ where $\widetilde{\nu }\left(
k\right) =\int \nu \left( X\right) e^{ikX}d^{4}X$ and the bias (\ref{m}), (%
\ref{b}) $N(k)$ becomes
\begin{equation*}
N(k)=1-\frac{4\pi ^{2}}{k^{2}}\left( \widetilde{\nu }\left( 0\right) -%
\widetilde{\nu }\left( k\right) \right) .
\end{equation*}%
In this limiting case by the choice of $\widetilde{\nu }\left( k\right) $,
we can assign $N(k)$ an arbitrary function of $k$. In other words we get
here a class of limiting topologies where Green functions $%
G_{true}=G_{0}\left( k\right) N\left( k\right) $ have a good ultraviolet
behavior and quantum field theories in such spaces turn out to be finite.

\subsection{Green function, general consideration}

The action (\ref{act}) remains invariant under translations $\vec{x}^{\prime
}=\vec{x}+\vec{c}$ with an arbitrary $\vec{c}$ which means that the measure (%
\ref{meas}) does not actually depend on the position of the center of mass
of the gas of wormholes and, therefore, we may restrict ourself with
homogeneous distributions $F(\xi )$ of wormholes in space only. Indeed, we
may define $d^{N}\xi =d^{N}\xi ^{\prime }d^{4}c$, while the integration over
$d^{4}c$ gives the volume of $R^{4}$ i.e., $\int $ $d^{4}c=L^{4}=V$ $\ $
which disappears from (\ref{meas}) due to the denominator \footnote{%
Technically, we may first restrict a portion of $R^{4}$ in the (\ref{act})
to a finite volume $V$ and then in final expressions consider the limit $%
V\rightarrow \infty $ (which represents the standard tool in thermodynamics
and QFT).}. In what follows we shall omit the prime from $\xi ^{\prime }$.

Let us consider the Fourier representation $N(x,x^{\prime },\xi )\rightarrow
N(k,k^{\prime },\xi )$ which in the case of a homogeneous distribution of
wormholes gives $N(k,k^{\prime })$ $=N(k,\xi )\delta (k-k^{\prime })$, then
we find
\begin{equation*}
G\left( k\right) =G_{0}\left( k\right) N(k,\xi )
\end{equation*}%
and the Green function can be taken as
\begin{equation}
G=\frac{N(k,\xi )}{k^{2}+m^{2}}.  \label{grf}
\end{equation}%
Then for the total partition function we find
\begin{equation}
Z_{total}\left( J\right) =\int \left[ DN(k)\right] e^{-I(N)}e^{-\frac{1}{2}%
\frac{L^{4}}{\left( 2\pi \right) ^{4}}\int \left( \frac{N(k)}{k^{2}+m^{2}}%
\left\vert J_{k}\right\vert ^{2}\right) dk},  \label{ztot}
\end{equation}%
where $\left[ DN\right] =\prod\limits_{k}dN_{k}$ and $\sigma (N)$ comes from
the integration measure (i.e., from the Jacobian of transformation from $%
F\left( \xi \right) $ to $N\left( k\right) $)
\begin{equation*}
e^{-I(N)}=\int \left[ DF\right] Z_{0}(N(k,\xi ))\delta \left( N\left(
k\right) -N\left( k,\xi \right) \right) .
\end{equation*}%
We point out that $I(N)$ can be changed by means of adding to the action (%
\ref{act} ) of an arbitrary "non-dynamical" constant term which depends only
on topology $S\rightarrow S+\Delta S(N\left( k\right) )$ (e.g., a
topological Euler term ). The multiplier $Z_{0}(N)$ defines the simplest
measure for topologies. Now by means of using the expression (\ref{grf}) and
(\ref{ztot}) we find the two-point Green function in the form
\begin{equation}
G\left( k\right) =\frac{\overline{N}(k)}{k^{2}+m^{2}}  \label{grfm}
\end{equation}%
where $\overline{N}(k)$ is the cutoff function (the mean bias) which is
given by \footnote{%
We recall that in this integral contribute only limiting topologies in which
density of wormholes diverges ($n\rightarrow \infty $).}
\begin{equation}
\overline{N}(k)=\frac{1}{Z_{total}\left( 0\right) }\int \left[ DN\right]
e^{-I\left( N\right) }N\left( k\right) .  \label{cutoff}
\end{equation}

At the present stage we still cannot evaluate the exact form for the cutoff
function $\overline{N}(k)$ in virtue of the ambiguity of $\Delta S(N\left(
k\right) )$ pointed out. Such a term may include two parts. First part $%
\Delta _{1}S$ describes the proper dynamics of wormholes and should be
considered separately. Indeed, in general wormholes are dynamical
self-gravitating objects which require considering the gravitational
contribution to the action. Some part of such a contribution (mean curvature
induced by wormholes) is discussed in Sec.6. However, since a wormhole
represents an extended non-local object, it possesses a rather complex
dynamics and this problem requires the further investigation. The second
part $\Delta _{2}S$ may describe "external conditions" (e.g., an external
classical field in (\ref{ztot})) for the mean topology. Actually the last
term can be used to prescribe an arbitrary particular value for the cutoff
function $\overline{N}(k)=f\left( k\right) $. Indeed, the "external
conditions" can be accounted for by adding the term $\Delta _{2}S=\left(
\lambda ,N\right) =\int \lambda \left( k\right) N\left( k\right) d^{4}k$,
where $\lambda \left( k\right) $ plays the role of a specific chemical
potential which implicitly depends on $f\left( k\right) $ through the
equation
\begin{equation}
f(k)=\frac{1}{Z_{total}\left( \lambda ,0\right) }\int \left[ DN\right]
e^{\left( \lambda ,N\right) -I\left( N\right) }N\left( k\right) .  \label{fc}
\end{equation}%
From (\ref{ztot}) we see that the role of such a chemical potential may play
the external current $\lambda \left( k\right) =-\frac{1}{2}\frac{L^{4}}{%
\left( 2\pi \right) ^{4}}G_{0}(k)\left\vert J_{k}^{ext}\right\vert
^{2}$ or equivalently an external classical field $\varphi
^{ext}=G_{0}(k)J_{k}^{ext}$. In quantum field theory such a term
leads merely to a renormalization of the cosmological constant. By
other words the mean topology (i.e., the cutoff function or mean
distribution of wormholes) is driven by the cosmological constant
$\Lambda $ and vise versa.

\section{Topological bias as a projection operator}

By the construction the topological bias $N\left( x,x^{\prime }\right) $
plays the role of a projection operator onto the space of functions (a
subspace of functions on $R^{4}$) which obey the proper boundary conditions
at throats of wormholes. This means that for any particular topology (for a
set of wormholes) there exists the basis $\{f_{i}\left( x\right) \}$ in
which it takes the diagonal form $N\left( x,x^{\prime }\right) =\sum
N_{i}f_{i}\left( x\right) f_{i}^{\ast }\left( x^{\prime }\right) $ with
eigenvalues $N_{i}=0,1$ (since $N_{i}^{2}=N_{i}$). In this section we
illustrate this simple fact (which is probably not obvious for readers) by
the explicit construction of the reference system for a single wormhole when
physical functions become (due to the boundary conditions) periodic
functions of one of coordinates.

Indeed, consider a single wormhole with parameters $\xi $ (i.e., $\xi
=\left( a,R^{+},R^{-}\right) $, where $a$ is the throat radius and $R^{\pm }$
are positions of throats in space \footnote{%
In general, there exists an additional parameter $\Lambda _{\beta }^{\alpha
} $ which defines a rotation of one of throats before gluing. However, it
does not change the subsequent construction. There always exists a
diffeomorphic map of coordinates $x^{\prime }=h\left( x\right) $ which sets
such a matrix to unity.}). Consider now a particular solution $\phi _{0}$ to
the equation $\Delta \phi _{0}=0$ (harmonic function) for $R^{4}$ in the
presence of the wormhole \footnote{%
Instead of the construction used here one may use also another method.
Indeed, consider two point charges, then the function $\phi _{0}=1/\left(
x-x_{+}\right) ^{2}-1/\left( x-x_{-}\right) ^{2}$ can be taken as a new
coordinate. Wormhole appears when we identify (glue) surfaces $\phi _{0}=\pm
\omega $. We point out that such surfaces are not spheres, though they
reduce to spheres in the limit $\left\vert x_{+}-x_{-}\right\vert
\rightarrow \infty $ or $\omega \rightarrow \infty $.}, which corresponds to
the situation when throats possess a unit charge/mass but those have the
opposite signs. Now define the family of lines of force $x\left(
s,x_{0}\right) $ which obey the equation $\frac{dx}{ds}=-\nabla \phi
_{0}\left( x\right) $ with initial conditions $x(0)=x_{0}$. Physically, such
lines correspond to lines of force for a two charged particles in positions $%
R^{\pm }$ with charges $\pm 1$. We note that all points which lay on the
trajectory $x\left( s,x_{0}\right) $ may be taken as initial conditions and
they define the same line of force with the obvious redefinition $%
s\rightarrow s-s_{0}$. By other words we may take as a new coordinates the
parameter $s$ and portion of the coordinates orthogonal to the family of
lines $x_{0}^{\perp }$. Coordinates $x_{0}^{\perp }$ can be taken as laying
in the hyperplane $R^{3}$ which is orthogonal to the vector $\vec{d}=\vec{R}%
^{-}-\vec{R}^{+}$ and goes through the point $\vec{X}_{0}=(\vec{R}^{-}+\vec{R%
}^{+})/2$.

Let $s^{\pm }\left( x_{0}^{\perp }\right) $ be the values of the parameter $%
s $ at which the line intersects the throats $R^{\pm }$. Then instead of $s$
we may consider a new parameter $\theta $ as $s\left( \theta \right)
=s^{-}+\left( s^{+}-s^{-}\right) \theta /2\pi $, so that when $\theta
=0,2\pi $ the parameter $s$ takes the values $s=s^{-},s^{+}$ respectively.
The gluing procedure at throats means merely that we identify points at $%
\theta =0$ and $\theta =2\pi $ and all physical functions in the space $%
R^{4} $ with a single wormhole $\xi $ become periodic functions of $\theta $%
. Thus, the coordinate transformation $x=x(\theta ,x_{0}^{\perp })$ gives
the map of the above space onto the cylinder with a specific metric $dl^{2}$
$=$ $\left( d\vec{x}\left( \theta ,x_{0}^{\perp }\right) \right) ^{2}$ $=$ $%
g_{\alpha \beta }dy^{\alpha }dy^{\beta }$ (where $y=(\theta ,x_{0}^{\perp })$%
) whose components are also periodic in terms of $\theta $. Now we can
continue the coordinates to the whole space $R^{4}$ (to construct a cover of
the fundamental region $\theta \in \lbrack 0,2\pi ]$) simply admitting all
values $-\infty <\theta <+\infty $ this, however, requires to introduce the
bias
\begin{equation*}
\frac{1}{\sqrt{g}}\delta \left( \theta -\theta ^{\prime }\right) \rightarrow
N\left( \theta -\theta ^{\prime }\right) =\sum_{n=-\infty }^{+\infty }\frac{1%
}{\sqrt{g}}\delta \left( \theta -\theta ^{\prime }+2\pi n\right) ,
\end{equation*}%
since every point and every source in the fundamental region acquires a
countable set of images in the non-physical region (inside of wormhole
throats). Considering now the Fourier transforms for $\theta $ we find
\begin{equation*}
N\left( k,k^{\prime }\right) =\sum_{n=-\infty }^{+\infty }\delta \left(
k-n\right) \delta \left( k-k^{\prime }\right) .
\end{equation*}%
We point out that the above bias gives the unit operator in the space of
periodic functions of $\theta $. From the standpoint of all possible
functions on $R^{4}$ it represents the projection operator $\widehat{N}^{2}=%
\widehat{N}\left( \xi \right) $ (taking an arbitrary function $f$ \ we find
that upon the projection $f_{N}=$\thinspace $\widehat{N}f$ $f_{N}$ becomes a
periodic function of $\theta $, i.e.,only periodic functions survive).

The above construction can be easily generalized to the presence
of a set of wormholes. In the approximation of a dilute gas of
wormholes we may neglect the influence of wormholes on each other
(at least there always exists a sufficiently smooth map which
transforms the family of lines of force for "independent"
wormholes onto the actual lines). Then the total bias (projection)
may be considered as the product
\begin{equation*}
N_{total}\left( x,x^{\prime }\right) =\int \left( \prod\limits_{i}\sqrt{g_{i}%
}d^{4}y_{i}\right) N\left( \xi _{1},x,y_{1}\right) N\left( \xi
_{2},y_{1},y_{2}\right) ...N\left( \xi _{N},y_{N-1},x^{\prime
}\right) ,
\end{equation*}%
where $N\left( \xi _{i},x,x^{\prime }\right) $ is the bias for a
single wormhole with parameters $\xi _{i}$. Every such a
particular bias $N\left( \xi _{i},x,x^{\prime }\right) $ realizes
projection on a subspace of functions which are periodic with
respect to a particular coordinate $\theta _{i}\left( x\right) $,
while the total bias gives the projection onto the intersection of
such particular subspaces (functions which are periodic with
respect to every parameter $\theta _{i}$).

\section{Cutoff}

The projective nature of the bias operator $N\left( x,x^{\prime }\right) $
allows us to express the cutoff function $\overline{N}(k)$ via dynamic
parameters of wormholes. Indeed, consider a box $L^{4}$ in $R^{4}$ and
periodic boundary conditions which gives $k=2\pi n/L$ (in final expressions
we consider the limit $L\rightarrow \infty $, which gives $%
\sum_{k}\rightarrow \frac{L^{4}}{\left( 2\pi \right) ^{4}}\int d^{4}k$). And
let us consider the decomposition for the integration measure in (\ref{ztot}%
) as
\begin{equation*}
I =I_{0}+\sum \lambda _{1}\left( k\right) N\left( k\right) +\frac{1}{2}\sum
\lambda _{2}\left( k,k^{\prime }\right) N\left( k\right) N\left( k^{\prime
}\right) +...
\end{equation*}%
where $\lambda _{1}\left( k\right) $ includes also the contribution from $%
Z_{0}\left( k\right) $. We point out that this measure plays the role of the
action for the bias $N\left( k\right) $. Indeed, the variation of the above
expression gives the equation of motions for the bias in the form
\begin{equation*}
\sum_{k^{\prime }}\lambda _{2}\left( k,k^{\prime }\right) N\left( k^{\prime
}\right) =-\lambda _{1}\left( k\right)
\end{equation*}%
which can be found by considering the proper dynamics of wormholes. We
however do not consider the problem of the dynamic description of wormholes
here and leave this for the future research. Then in the first approximation
we may retain the linear term only \footnote{%
Indeed, the projective nature of the bias $N\left( k\right) =0,1$ means that
it can be phenomenologically expressed via some Fermionic ghost field $\Psi
\left( k\right) $ (e.g., $N\left( k,k^{\prime }\right) =\Psi \left( k\right)
\Psi ^{+}\left( k^{\prime }\right) $) where the negative and positive
frequency parts of the operator $\Psi \left( k\right) $ obey the
anti-commutation relations $\Psi ^{+}\left( k\right) \Psi \left( k^{\prime
}\right) +\Psi \left( k^{\prime }\right) \Psi ^{+}\left( k\right) =\delta
\left( k-k^{\prime }\right) $. In the absence of ghost particles $\Psi
\left( k\right) \left\vert 0\right\rangle =0$ we get $N\left( k,k^{\prime
}\right) =\delta \left( k-k^{\prime }\right) $, i.e., $N\left( k\right) =1$
and wormholes are absent. In the term of the ghost field the action becomes $%
I\left( N\right) =I_{0}+\left( \Psi ,\widehat{\lambda }_{1}\Psi \right) +...$%
. Therefore in the leading approximation equations of motion take the linear
form $\widehat{\lambda }_{1}\Psi =0$.}. Then taken into account that $%
N\left( k\right) =0,1$ ($N^{2}=N$) we find
\begin{equation}
\overline{N}(k)=\frac{1}{Z_{total}\left( k\right) }\sum_{N=0,1}e^{-\lambda
_{1}\left( k\right) N\left( k\right) }N\left( k\right) =\frac{e^{-\lambda
_{1}\left( k\right) }}{1+e^{-\lambda _{1}\left( k\right) }}.  \label{mn}
\end{equation}

The simplest choice gives merely $\lambda _{1}\left( k\right) =-\sum \ln
Z_{0}\left( k\right) $, where the sum is taken over the number of fields and
$Z_{0}\left( k\right) $ is given by $Z_{0}\left( k\right) =\sqrt{\pi
/(k^{2}+m^{2})}$. In the case of a set of massless fields we find $\overline{%
N}(k)=Z\left( k\right) /\left( 1+Z\left( k\right) \right) $ where $Z\left(
k\right) =\left( \sqrt{\pi }/k\right) ^{\alpha }$ and $\alpha $ is the
effective number of degrees of freedom (the number of boson minus fermion
fields). To ensure the absence of divergencies one has to consider the
number of fields $\alpha >4$ \cite{KS09}. However, such a choice gives the
simplest estimate which, in general, cannot be correct. Indeed, while its
behavior at very small scales (i.e., when exceeding the Plankian scales $%
Z\left( k\right) \lesssim 1$ and $\overline{N}(k)=Z\left( k\right) $) may be
physically accepted, since it produces some kind of a cutoff, on the
mass-shell $k^{2}+m^{2}\rightarrow 0$ it gives the behavior $\overline{N}%
(k)\rightarrow 1$ which is merely incorrect (e.g., from (\ref{mnl}) we see
that the true behavior should be $\overline{N}(k)\rightarrow const<1$).

One may expect that the true cutoff function has a much more complex
behavior. Indeed, some theoretical models in particle physics (e.g., string
theory) have the property to be lower-dimensional at very small scales. The
mean cutoff $\overline{N}(k)$ gives the natural tool to describe a
scale-dependant dimensional reduction \cite{K03}. In fact, this function
defines the spectral number of modes in the interval between $k$ and $k+dk$
as
\begin{equation*}
\int \overline{N}(k)\frac{d^{4}k}{(2\pi )^{4}}=\int \frac{\overline{N}%
(k)k^{4}}{(2\pi )^{2}}\frac{dk}{k}.
\end{equation*}%
Hence we can define the effective spectral dimension $D$ of space as follows
\begin{equation*}
k^{4}\overline{N}(k)\sim k^{D}.
\end{equation*}%
From the empirical standpoint the dimension $D=4$ is verified at laboratory
scales only, while the rigorous tool to define the spectral \ density of
states (or the mean cutoff) can give the lattice quantum gravity e.g., see
\cite{AJL,MRS} and references therein. And indeed, the spectral dimension
for nonperturbative quantum gravity defined via Euclidean dynamical
triangulations was calculated recently in \cite{LQG}. It turns out that it
runs from a value of $D=3/2$ at short distance to $D=4$ at large distance
scales. We also point out that all observed dark matter phenomena can be
explained by the fractal dimension $D\approx 2$ starting from scales $%
L\gtrsim (1\div 5)Kpc$, e.g., \cite{KT02}.

\section{Cosmological constant}

Let us consider the total Euclidean action \cite{H78}%
\begin{equation}
I_{E}=-\frac{1}{16\pi G}\int (R-2\Lambda _{0})\sqrt{g}d^{4}x-\int L_{m}\sqrt{%
g}d^{4}x.
\end{equation}%
The variation of the above action leads to the Einstein equations%
\begin{equation*}
R_{ab}-\frac{1}{2}g_{ab}R+g_{ab}\Lambda _{0}=8\pi GT_{ab}
\end{equation*}%
where $T^{ab}=\frac{1}{2}(g)^{-1/2}(\delta L_{m}/\delta g_{ab})$ is the
stress energy tensor and $\Lambda _{0}$ is a naked cosmological constant. In
cosmology such equations are considered from the classical standpoint ,
which means that they involve characteristic scales $\ell \gg \ell _{pl}$.
However, the presence of virtual wormholes at planckian scales defines some
additional contribution in both parts of these equations which can be
adsorbed into the cosmological constant. Therefore the total cosmological
constant can be defined as%
\begin{equation*}
\Lambda _{tot}=\Lambda _{0}+\Lambda _{m}+\Lambda _{R}=\Lambda _{0}+2\pi G<T>+%
\frac{1}{4}<R>_{w}
\end{equation*}%
where $<T>$ is the energy of zero-point fluctuations \footnote{%
It includes also the contribution of zero-point fluctuations of gravitons.}
i.e., the mean vacuum value (we recall that in the standard QFT $\Lambda
_{m} $ is infinite, while wormholes form a finite value) and $%
<R>_{w}=\Lambda _{R} $ is a contribution of wormholes into the mean
curvature due to gluing (\ref{wmetr}).

\subsection{Contribution of virtual wormholes into mean curvature}

Consider a single wormhole whose metric is given by (\ref{wmetr}) $%
ds^{2}=h^{2}\left( r\right) \delta _{\alpha \beta }dx^{\alpha }dx^{\beta }$,
then the components of the Ricci tensor are \footnote{%
We are much obliged to the Referee who pointed out to the subtleties when
working with a step function in the metric.}%
\begin{equation}\label{Ricci}
R_{\alpha \beta }=-\left( \frac{h^{\prime }}{h}\right) ^{\prime
}\left( \delta _{\alpha \beta }+(N-2)n_{\alpha }n_{\beta }\right)
-
\frac{1}{r}\frac{%
h^{\prime }}{h}\left( \left( N-2\right) \Delta _{\alpha \beta }+
\delta _{\alpha \beta }(N-1)\right) -\left( \frac{h^{\prime
}}{h}\right) ^{2}\left( N-2\right) \Delta _{\alpha \beta }
\end{equation}%
where $n^{\nu }=x^{\nu }/a$ is the unite normal vector to the
throat surface, $\Delta _{\alpha \beta }=\delta _{\alpha \beta
}-n_{\alpha }n_{\beta }$, $h^{\prime }=\partial h/\partial r$, and
$N$ is the number of
dimensions. The curvature scalar is%
\begin{equation}
R=-\frac{\left( N-1\right) }{h^{2}}\left[ 2\frac{h^{\prime \prime }}{h}+2%
\frac{1}{r}\frac{h^{\prime }}{h}(N-1)+\left( \frac{h^{\prime }}{h}\right)
^{2}\left( N-4\right) \right]
\end{equation}%
Then substituting $h=1+\left( \frac{a^{2}}{r^{2}}-1\right) \theta $ in the
above equation we find

\begin{equation}
-\frac{h^{4}}{N-1}R=4h\frac{a^{2}}{r^{3}}\delta +\frac{2h}{r^{N-1}}\left(
r^{N-1}\lambda \right) ^{\prime }+\left( N-4\right) \left( \lambda
^{2}-4\theta \frac{a^{2}}{r^{3}}\lambda \right) +4\left( N-4\right) \frac{%
a^{2}}{r^{4}}\theta \left( \theta -1\right)  \label{R}
\end{equation}%
where $\theta $ is a smooth function which only in a limit becomes a step
function, $\lambda =\left( 1-\frac{a^{2}}{r^{2}}\right) \delta $, and $%
\delta =-\theta ^{\prime }$. In the limit $\theta \rightarrow \theta \left(
a-r\right) $ we find $\delta \rightarrow \delta (r-a)$ and the last two
terms in (\ref{R}) are negligible as compared to the first two terms, while
in four dimensions the last two terms vanish. Then the curvature is
concentrated on the throat of the wormhole where $\theta $ differs from the
step function. In the limit of a vanishing throat size we have $h\rightarrow
1$ at the throat and we get

\begin{equation}
-R=\frac{4\left( N-1\right) }{a}\delta \left( r-a\right) +\frac{2\left(
N-1\right) }{r^{N-1}}\left( r^{N-3}\left( r^{2}-a^{2}\right) \delta \right)
^{\prime }.
\end{equation}%
In general all the terms together in (\ref{R}) and analogous terms in the
Ricci tensor provide that the Bianchi identity holds and the energy is
conserved \cite{taub}. Therefore none of them can be dropped out. However it
is easy to see the leading contribution to the integral over space comes
from the first term only and in the limit of a vanishing throat size we find
for a single wormhole%
\begin{equation}
\frac{1}{4}\int R\sqrt{g}d^{4}x=-6\pi ^{2}a^{2}.
\end{equation}

In the case of a set of wormholes (\ref{ff}) we find
\begin{equation*}
\frac{1}{4}\int R\sqrt{g}d^{4}x=-6\pi ^{2}\sum_{j}a_{j}^{2}=-12\pi ^{2}\int
n\left( a\right) a^{2}dad^{4}x=\int \Lambda _{R}d^{4}x
\end{equation*}%
where $2n\left( a\right) =\int n\left( a,r_{0}\right) dr_{0}$ is the density
of wormhole throats with a fixed value of the throat size $a$. This defines
the contribution to the cosmological constant from the mean curvature as
\begin{equation}
\Lambda _{R}=-12\pi ^{2}\int n\left( a\right) a^{2}da<0.  \label{mcv}
\end{equation}%
We see that this quantity is always negative.

\subsection{Stress energy tensor}

In this section we consider the contribution from matter fields. In the case
of a scalar field the stress energy tensor has the form%
\begin{equation*}
-T_{\alpha \beta }\left( x\right) =\partial _{\alpha }\varphi \partial
_{\beta }\varphi -\frac{1}{2}g_{\alpha \beta }\left( \partial ^{\mu }\varphi
\partial _{\mu }\varphi +m^{2}\varphi ^{2}\right) .
\end{equation*}%
Then the mean vacuum value of the stress energy tensor can be obtained
directly from the two-point green function (\ref{grf}), (\ref{grfm}) as
\begin{equation}
-\left\langle T_{\alpha \beta }\left( x\right) \right\rangle
=\lim_{x^{\prime }\rightarrow x}\left( \partial _{\alpha }\partial _{\beta
}^{\prime }-\frac{1}{2}g_{\alpha \beta }\left( \partial ^{\mu }\partial
_{\mu }^{\prime }+m^{2}\right) \right) \left\langle G\left( x,x^{\prime
},\xi \right) \right\rangle .  \label{t0}
\end{equation}%
By means of using the Fourier transform $G\left( x,x^{\prime },\xi \right)
=\int e^{-ik(x-x^{\prime })}G\left( k,\xi \right) \frac{d^{4}k}{\left( 2\pi
\right) ^{4}}$ and the expression (\ref{grf}), (\ref{grfm}) we arrive at
\begin{equation}
\left\langle T_{\alpha \beta }\left( x\right) \right\rangle =\frac{1}{4}%
g_{\alpha \beta }\int \left( 1+\frac{m^{2}}{k^{2}+m^{2}}\right) \overline{N}%
(k)\frac{d^{4}k}{\left( 2\pi \right) ^{4}},  \label{t}
\end{equation}%
where the property $\int k_{\alpha }k_{\beta }f\left( k^{2}\right) d^{4}k=$ $%
\frac{1}{4}g_{\alpha \beta }\int k^{2}f\left( k^{2}\right) d^{4}k$ has been
used and $\overline{N}(k)=\left\langle N\left( k,\xi \right) \right\rangle $
is the cutoff function (\ref{cutoff}).

For the sake of simplicity we consider the massless case. Then by the use of
the cutoff $\overline{N}(k)=\pi ^{\alpha /2}/\left( \pi ^{\alpha
/2}+k^{\alpha }\right) $ from the previous section we get the finite
estimate ($\alpha >4$ is the effective number of the field helicity states)
\begin{equation}
\Lambda _{m}=2\pi G\sum^{\alpha }\int \frac{\pi ^{\alpha /2}}{\pi ^{\alpha
/2}+k^{\alpha }}\frac{d^{4}k}{\left( 2\pi \right) ^{4}}=\frac{\pi G}{4}%
\Gamma \left( \frac{\alpha -4}{\alpha }\right) \Gamma \left( \frac{4}{\alpha
}\right) \sim 1,  \label{cc}
\end{equation}%
where the sum is taken over the number of fields. Since the leading
contribution comes here from very small scales, we may hope that this value
will not essentially change if the true cutoff function changes the behavior
on the mass-shell as $k\rightarrow 0$ (e.g., if we take $\lambda _{1}\left(
k\right) =-\sum \ln Z_{0}\left( k\right) +\delta \lambda \left( k\right) $
with $\delta \lambda \left( k\right) \ll \ln Z_{0}\left( k\right) $ as $k\gg
1$).

To understand how wormholes remove divergencies, it will be convenient to
split the bias function into two parts $N\left( k,\xi \right) =1+b\left(
k,\xi \right) $, where $1$ corresponds to the standard Euclidean
contribution, while $b\left( k,\xi \right) $ is the contribution of
wormholes. The first part gives the well-known divergent contribution of
vacuum field fluctuations $8\pi G\left\langle T_{\alpha \beta
}^{0}\right\rangle =\Lambda _{\ast }g_{\alpha \beta }$ with $\Lambda _{\ast
}\rightarrow +\infty \,$, while the second part remains finite for any
finite number of wormholes and, due to the projective nature of the bias
described in the previous section, it partially compensates (reduces) the
value of the cosmological constant, i.e., $8\pi G\left\langle \Delta
T_{\alpha \beta }\right\rangle =\delta \Lambda g_{\alpha \beta }$, where $%
\delta \Lambda =\sum_{N}\rho _{N}\delta \Lambda \left( N\right) $ and $%
\delta \Lambda \left( N\right) $ is a negative finite contribution of a
finite set of wormholes.

Consider now the particular distribution of virtual wormholes (\ref{ff}) and
evaluate their contribution to the cosmological constant which is given by $%
\delta \Lambda \left( N\right) $ $=$ $2\pi G\int b\left( k\right) \frac{%
d^{4}k}{\left( 2\pi \right) ^{4}}$ $=2\pi Gb_{total}\left( 0\right) $. Then
from the expressions (\ref{bt}) and (\ref{ff}) we get
\begin{equation*}
b_{total}\left( 0\right) =-\frac{n}{4\pi ^{4}a^{3}r_{0}^{3}}\int \left( 1-%
\frac{a^{2}}{R_{-}^{2}}\right) \delta (R_{+}-a)\delta \left( \left\vert
R_{+}-R_{-}\right\vert -r_{0}\right) d^{4}R_{-}d^{4}R_{+}
\end{equation*}%
which gives%
\begin{equation*}
b_{total}\left( 0\right) =-n\left( 1-f\left( \frac{a}{r_{0}}\right) \right) ,
\end{equation*}%
where
\begin{equation*}
f\left( \frac{a}{r_{0}}\right) =\frac{2}{\pi }\int_{0}^{\pi }\frac{a^{2}\sin
^{2}\theta d\theta }{a^{2}+2ar_{0}\cos \theta +r_{0}^{2}}.
\end{equation*}%
For $a/r_{0}\ll 1$ (we recall that by the construction $a/r_{0}\leq 1/2$)
this function has the value $f\left( \frac{a}{r_{0}}\right) \approx
a^{2}/r_{0}^{2}$. Thus, for the contribution of wormholes we find
\begin{equation*}
\delta \Lambda _{m}=-2\pi G\int n\left( a,r_{0}\right) \left( 1-f\left(
\frac{a}{r_{0}}\right) \right) dadr_{0}=-2\pi Gn\left( 1-\left\langle
f\right\rangle \right) .
\end{equation*}

\subsection{Vacuum value of the cosmological constant}

From the above expression we see that to get the finite value of the
cosmological constant $\Lambda _{m}=\Lambda _{\ast }+\delta \Lambda
_{m}<\infty $ one should consider the limit $n\rightarrow \infty $ (infinite
density of virtual wormholes) which requires considering the smaller and
smaller wormholes. From the other hand we have the obvious restriction $\int
2n(a,r_{0})\frac{\pi ^{2}}{2}a^{4}dadr_{0}<1$, where $\frac{\pi ^{2}}{2}%
a^{4} $ is the volume of one throat (wormholes cannot cut more, than the
total volume of space) \footnote{%
We also point out that in removing divergencies the leading role plays the
zero-point energy. Indeed $\delta \Lambda _{m}\sim -2\pi Gn$, while the mean
curvature has the order $\Lambda _{R}\sim -a^{2}n$ and for $a\ll \ell _{pl}$
we have $\delta \Lambda _{m}$ $\gg \Lambda _{R}$. Moreover in the limit $%
n\rightarrow \infty $, one gets $a\rightarrow 0$ and therefore $\Lambda
_{R}/\delta \Lambda _{m}\rightarrow 0$.}. Therefore, in the leading order it
seems to be sufficient to retain point-like wormholes only (i.e., consider
the limit $a\rightarrow 0$). Then instead of (\ref{ff}) we may assume the
vacuum distribution of virtual wormholes in the form%
\begin{equation*}
NF\left( a,X\right) =\frac{1}{a^{2}}\delta \left( a\right) \nu \left(
X\right) ,
\end{equation*}%
where $\nu \left( X\right) =\int a^{2}NF\left( a,X\right) da$ and $\int
\frac{1}{a^{2}}\nu \left( X\right) d^{4}X=n\rightarrow \infty $ has the
meaning of the infinite density of point-like wormholes, while $\nu \sim
a^{2}n$ remains a finite. In this case the volume cut by wormholes vanishes $%
\int 2n\frac{\pi ^{2}}{2}a^{4}dadr_{0}=a^{2}\int \nu \left( X\right)
d^{4}X\rightarrow 0$ and the rarefied gas approximation works well. This
defines the bias and the mean cutoff (here we define the Fourier transform $%
\widetilde{\nu }\left( k\right) =\int \nu \left( X\right) e^{ikX}d^{4}X$) as
\begin{equation*}
\overline{N}(k)=1-\frac{4\pi ^{2}}{k^{2}}\left( \widetilde{\nu }\left(
0\right) -\widetilde{\nu }\left( k\right) \right) .
\end{equation*}%
The contribution to the mean curvature (\ref{mcv}) can be expressed via the
same function $\nu \left( k\right) $ as
\begin{equation*}
\Lambda _{R}=-12\pi ^{2}\int n\left( a,r_{0}\right) a^{2}dadr_{0}=-12\pi
^{2}\int \nu \left( X\right) d^{4}X=-12\pi ^{2}\widetilde{\nu }\left(
0\right) .
\end{equation*}%
Thus, for the total cosmological constant we get the expression%
\begin{equation}
\Lambda _{tot}=\Lambda _{0}+2\pi G\int \left( 1-\frac{4\pi ^{2}}{k^{2}}%
\left( \widetilde{\nu }\left( 0\right) -\widetilde{\nu }\left( k\right)
\right) \right) \frac{d^{4}k}{\left( 2\pi \right) ^{4}}-12\pi ^{2}\widetilde{%
\nu }\left( 0\right) .  \label{le}
\end{equation}%
We stress that all these terms should be finite. Indeed all distributions of
virtual wormholes $\widetilde{\nu }\left( k\right) $ which lead to an
infinite value of $\Lambda _{tot}$ are suppressed in (\ref{z1}) by the
factor $\sim e^{-\int \Lambda _{tot}d^{4}x}$, while the minimal value is
reached when wormholes cut all of the volume of space and the action \ is
merely $S=0$ \footnote{%
Frankly speaking this statement is not rigorous. At first look the two last
terms in (\ref{le}) are independent and one may try to take $\widetilde{\nu }%
(0)$ an arbitrary big. If this were the case then the action would not
possess the minimum at all. However $\widetilde{\nu }(0)$ cannot be
arbitrary big, since it will violate the rarefied gas approximation and the
linear expression (\ref{le}) brakes down. Moreover, fermions give here a
contribution of the opposite sign. The rigorous investigation of this
problem requires the further studying and we present it elsewhere.}. We also
point out that here we considered the real scalar field as the matter
source, while in the general case the stress energy tensor should include
all existing bose and fermi fields (fermi fields give a negative
contribution to $\Lambda _{m}$).

The value of $\Lambda _{0}$ looks like a free parameter, which in quantum
gravity \ runs with scales \cite{MRS}. However at large scales its
asymptotic value may be uniquely fixed by the simple arguments as follows.
Indeed in quantum field theory properties of the ground state (vacuum)
change when we imply an external classical fields. The same is true for the
distribution of virtual wormholes $\widetilde{\nu }\left( k,J\right) $ e.g.,
see \ (\ref{ztot}) and (\ref{fc}) and, therefore, $\Lambda _{tot}=\Lambda
_{tot}(J)$ which we describe in the next subsection. We recall that in
gravity the role of the external current plays the stress energy tensor of
matter fields $J=T_{ab}$. However one believes that in the absence of all
classical fields the vacuum state should represent the most symmetric
(Lorentz invariant) state which in our case corresponds to the Euclidean
space. In order to be consistent with the Einstein equations this requires $%
\Lambda _{tot}(J=0)=0$ which uniquely fixes the value of $\Lambda _{0}$ in (%
\ref{le}). We point out that from somewhat different considerations such a
choice was advocated earlier in \cite{col,hwu}.

\subsection{Vacuum polarization in an external field}

Consider now topology fluctuations in the presence of an external current.
In the presence of an external current $J^{ext}$ the distribution of virtual
wormholes changes $\widetilde{\nu }\left( k,J\right) =\widetilde{\nu }\left(
k\right) +\delta \widetilde{\nu }\left( k,J\right) $. Indeed in (\ref{ztot})
for the case of a weak external field the contribution of the external
current into the action can be expanded as $\exp (-V(J))\simeq 1-V$, where
\begin{equation*}
V=-\frac{1}{2}\int J(x)G(x,y)J(y))d^{4}xd^{4}y=-\frac{1}{2}\frac{L^{4}}{%
\left( 2\pi \right) ^{4}}\int G_{0}(k)N(k)\left\vert J_{k}\right\vert
^{2}d^{4}k.
\end{equation*}%
Then using (\ref{cutoff}) we find $\overline{N}(k,J)=\overline{N}%
(k,0)+\delta N(k,J)$, where
\begin{equation*}
\delta N(k,J)=\delta b\left( J\right) \simeq -\frac{1}{2}\frac{L^{4}}{\left(
2\pi \right) ^{4}}\int \sigma ^{2}(k,p)G_{0}(p)\left\vert J_{p}\right\vert
^{2}d^{4}p
\end{equation*}%
is the bias related to an additional distribution of virtual wormholes and $%
\sigma ^{2}(k,p)=\overline{\Delta N^{\ast }(k)\Delta N(p)}$
\begin{equation*}
\sigma ^{2}(k,p)=\frac{1}{Z_{total}\left( 0\right) }\int \left[ DN\right]
e^{-I\left( N\right) }\Delta N^{\ast }\left( k\right) \Delta N(p)
\end{equation*}%
defines the dispersion of vacuum topology fluctuations (here $\Delta N=N-%
\overline{N}$). The exact definition of $\sigma ^{2}(k,p)$ requires the
further development of a fundamental theory. In particular, it can be
numerically calculated in lattice quantum gravity \cite{LQG}. However it can
be shown that at scales $k,p\gg k_{pl}$ it reduces to $\sigma
^{2}(k,p)\rightarrow \sigma _{k}^{2}\delta (k-p)$ and therefore
\begin{equation}
\delta b\left( J\right) =-\sigma _{k}^{2}\frac{4\pi ^{2}}{2k^{2}}\left\vert
J_{k}^{ext}\right\vert ^{2}.
\end{equation}%
Now comparing this function with (\ref{bk}) we relate the additional
distribution of virtual wormholes and the external classical field as%
\begin{equation}
\frac{4\pi ^{2}}{k^{2}}\left( \delta \widetilde{\nu }\left( 0,J\right)
-\delta \widetilde{\nu }\left( k,J\right) \right) =\frac{1}{2}\sigma _{k}^{2}%
\frac{4\pi ^{2}}{k^{2}}\left\vert J_{k}^{ext}\right\vert ^{2},  \label{fl}
\end{equation}%
where $\delta \widetilde{\nu }\left( k,J\right) =\int a^{2}\delta NF\left(
a,k\right) da$. We point out that the above expression does not define the
value $\delta \widetilde{\nu }\left( 0,J\right) $ which requires an
additional consideration. Moreover, in general the external field $J$ does
not possesses a symmetry and therefore the correction $\left\langle \delta
T_{\alpha \beta }\left( x\right) \right\rangle $ does not reduce to a single
cosmological constant. However, such corrections always violate the averaged
null energy condition \cite{GR} \ and may be considered as some kind of dark
energy or, by other words, it represents an exotic matter. Some portion of
dark energy still has the form of the cosmological constant which defines a
non-vanishing present day value (we recall that fermions give a contribution
of the opposite sign)%
\begin{equation}
\delta \Lambda _{tot}=-2\pi G\int \frac{4\pi ^{2}}{k^{2}}\left\langle \delta
\widetilde{\nu }\left( 0,J\right) -\delta \widetilde{\nu }\left( k,J\right)
\right\rangle \frac{d^{4}k}{\left( 2\pi \right) ^{4}}-12\pi ^{2}\delta
\widetilde{\nu }\left( 0,J\right) ,
\end{equation}%
where $\left\langle \delta \widetilde{\nu }\left( k,J\right) \right\rangle $
denotes an averaging over rotations.

The only unknown parameter in (\ref{fl}) is the dispersion $\sigma _{k}^{2}$
which defines the intensity of topology fluctuations in the vacuum. It has
also the sense of the efficiency coefficient which defines the portion of
the energy of the external field spent on the formation of additional
wormholes. Though the evaluation of $\sigma _{k}^{2}$ requires the further
development of a fundamental theory, one may expect that $\sigma _{k}^{2}=%
\overline{N}(k)(1-\overline{N}(k))$ where $\overline{N}(k)$ is the mean
cutoff. It is expected that $\overline{N}\left( k\right) \rightarrow 0$ as $%
k\gg k_{pl}$ and $\overline{N}\left( k\right) \rightarrow \overline{N}%
_{0}\leq 1$. This means that $\sigma \rightarrow 0$ as $k\gg k_{pl}$ and $%
\sigma \rightarrow \sigma _{0}\ll 1$ as $k\ll k_{pl}$, while it takes the
maximum value $\sigma _{\max }\sim 1$ at Planckian scales $k \sim k_{pl}$.
By other words, the most efficient transmission of the energy into wormholes
takes place for wormholes of the Planckian size. \ In the case when external
classical fields have characteristic scales $\lambda =2\pi /k\gg \ell _{pl}$
in (\ref{fl}) the efficiency coefficient $\sigma _{k}^{2}$ and the cutoff $%
\overline{N}\left( k\right) $ become constant $\sigma \simeq \sigma _{0}$, $%
\overline{N}\left( k\right) \simeq \overline{N}_{0}$, while their ratio may
be estimated as $\alpha \sigma _{0}^{2}/\overline{N}_{0}=\Omega _{DE}/\Omega
_{b}$, where $\alpha $ is the effective number of fundamental fields which
contribute to $\delta \Lambda _{tot}$ and $\Omega _{DE}$, $\Omega _{b}$ are
dark energy and baryon energy densities respectively. According to the
modern picture this ratio gives $\Omega _{DE}/\Omega _{b}\approx
0.75/0.05=15 $, while $\sigma _{0}^{2}/\overline{N}_{0}\sim 1$ (as $%
\overline{N}_{0}\ll 1$) and therefore this defines the estimate for the
effective number of fundamental fields (helicity states) as $\alpha \sim 15$.

\subsection{Speculations on the formation of actual wormholes}

As we already pointed out the additional distribution of virtual wormholes (%
\ref{fl}) reflects the symmetry of external classical fields and therefore
it forms a homogeneous and isotropic background and perturbations. We recall
that virtual wormholes represent an exotic form of matter. In the early
Universe such perturbations start to develop and may form actual wormholes.
The rigorous description of such a process represents an extremely complex
and interesting problem which requires the further study. Some aspects of
the behavior of the exotic density perturbations were considered in \cite%
{KS10}, while the simplest example of the formation of a wormhole-type
object was discussed recently by us in \cite{KS12}. Therefore we may expect
that some portion of such a form of dark energy is reserved now in actual
wormholes which we consider in the next section.

\section{Dark energy from actual wormholes}

Consider now the contribution to the dark energy from the gas of actual
wormholes. Unlike the virtual wormholes, actual wormholes do exist at all
times and, therefore, a single wormhole can be viewed as a couple of
conjugated cylinders $T_{\pm }^{3}=S_{\pm }^{2}\times R^{1}$. So that the
number of parameters of an actual wormhole is less $\eta =(a,r_{+},r_{-})$,
where $a$ is the radius of $S_{\pm }^{2}$ and $r_{\pm }$ $\in R^{3}$ is a
spatial part of $R_{\pm }$.

Actual wormholes also produce two kind of contribution to the dark energy.
One comes from their contribution to the mean curvature which correspond to
an exotic stress energy momentum tensor. Such a stress energy momentum
tensor reflects the dark energy reserved by additional virtual wormholes
discussed in the previous section. Such energy is necessary to support
actual wormholes as a solution to the Einstein equations. The second part
comes from vacuum polarization effects by actual wormholes. The
consideration in the previous section shows that for macroscopic wormholes
the second part has the order $\left\langle \Delta T_{\alpha \beta
}\right\rangle \sim 8\pi Gn$ and is negligible as compared to the curvature $%
R\sim a^{2}n$ (since macroscopic wormholes have throats $a\gg \ell _{pl}$).
However, for the sake of completeness and for methodological aims we
describe it as well.

For rigorous evaluation of dark energy of the second type we, first, have to
find the bias $b_{1}\left( x,x^{\prime },\eta \right) $ analogous to (\ref%
{b1}) for the topology $R^{4}/(T_{+}^{3}\cup T_{-}^{3})$. There are many
papers treating different wormholes in this respect (e.g., see \cite{GR} and
references therein). However, in the present paper for an estimation we
shall use a more simple trick.

\subsection{Beads of virtual wormholes (quantum wormhole)}

Indeed, instead of the cylinders $T_{\pm }^{3}$ we consider a couple of
chains (beads of virtual wormholes $T_{\pm }^{3}\rightarrow \cup _{n}S_{\pm
,n}^{3}$). Such an idea was first suggested in \cite{M77} and one may call
such an object as quantum wormhole. Then the bias can be written
straightforwardly
\begin{equation}
b_{1}\left( x,x^{\prime },\eta \right) =\sum_{n=-\infty }^{+\infty }\frac{1}{%
4\pi ^{2}a}\left( \frac{1}{\left( R_{-,n}-x^{\prime }\right) ^{2}}-\frac{1}{%
\left( R_{+,n}-x^{\prime }\right) ^{2}}\right) \times  \label{brw}
\end{equation}%
\begin{equation*}
\times \left[ \delta (\left\vert \vec{x}-\vec{R}_{+,n}\right\vert -a)-\delta
(\left\vert \vec{x}-\vec{R}_{-,n}\right\vert -a)\right] ,
\end{equation*}%
where $R_{\pm ,n}=\left( t_{n},r_{\pm }\right) $ with $t_{n}=t_{0}+2\ell n$
and $\ell \geq a$ is the step. We may expect that upon averaging over the
position $t_{0}\in \lbrack -\ell ,\ell ]$ the bias for the beads will
reproduce the bias for cylinders $T_{\pm }^{3}$ ( at least it looks like a
very good approximation). We point out that the averaging out $\frac{1}{%
2\ell }\int_{-\ell }^{\ell }dt_{0}$ and the sum $\sum_{n=-\infty }^{+\infty
} $ reduces to a single integral $\frac{1}{2\ell }\int_{-\infty }^{\infty
}dt $ of the zero term in (\ref{brw}). And moreover, the resulting total
bias corresponds merely to a specific choice of the distribution function $%
F(\xi ) $ in (\ref{bs}). Namely, we may take
\begin{equation*}
NF(\xi )=\frac{1}{2\ell }\delta \left( t_{+}-t_{-}\right) f\left( \left\vert
r_{+}-r_{-}\right\vert ,a\right) ,
\end{equation*}%
where $R_{\pm }=\left( t_{\pm },r_{\pm }\right) $ and $f\left( s,a\right) $
is the distribution of cylinders, which can be taken as ($\widetilde{n}$ is
3-dimensional density)
\begin{equation*}
f\left( \eta \right) =\frac{\widetilde{n}\left( a\right) }{4\pi r_{0}^{2}}%
\delta \left( s-r_{0}\right) .
\end{equation*}%
Using the normalization condition $\int NF(\xi )d\xi =N$ we find the
relation $N=\frac{1}{2\ell }\widetilde{n}V=$ $nV$, where $n$ is a
4-dimensional density of wormholes and $1/\left( 2\ell \right) $ is the
effective number of wormholes on the unit length of the cylinder (i.e., the
frequency with which the virtual wormhole appears at the positions $r_{\pm }$
). This frequency is uniquely fixed by the requirement that the volume which
cuts the bead is equal to that which cuts the cylinder $\frac{4}{3}\pi a^{3}=%
\frac{\pi ^{2}}{2}a^{4}\frac{1}{2\ell }$ (i.e., $2\ell =\frac{3\pi }{8}a$
and $n=\frac{8}{3\pi a}\widetilde{n}$). Thus, we can use directly expression
(\ref{bk}) and find (compare to (\ref{b(k)})
\begin{equation}
b\left( k\right) =-\int n\left( a\right) a^{2}\frac{4\pi ^{2}}{k^{2}}\left(
1-\frac{\sin \left\vert \mathbf{k}\right\vert r_{0}}{\left\vert \mathbf{k}%
\right\vert r_{0}}\right) \frac{J_{1}\left( ka\right) }{ka/2}da,  \label{bd}
\end{equation}%
where $k=(k_{0},\mathbf{k})$. Here the first term merely coincides with that
in (\ref{b(k)}) and, therefore, it gives the contribution to the
cosmological constant $\delta \Lambda /(8\pi G)=-n/4=-2\widetilde{n}/\left(
3\pi a\right) $, while the second term describes a correction which does not
reduce to the cosmological constant and requires a separate consideration.

\subsection{Stress energy tensor}

From (\ref{t0}) we find that the stress energy tensor
\begin{equation}
-\left\langle \Delta T_{\alpha \beta }\left( x\right) \right\rangle =\int
\frac{k_{\beta }k_{\alpha }-\frac{1}{2}g_{\alpha \beta }k^{2}}{k^{2}}b\left(
k,\xi \right) \frac{d^{4}k}{\left( 2\pi \right) ^{4}}
\end{equation}%
reduces to the two functions
\begin{equation*}
T_{00}=\varepsilon =\lambda _{1}-\frac{1}{2}\mu
\end{equation*}%
\begin{equation*}
T_{ij}=p\delta _{ij},~~p=\frac{1}{3}\lambda _{2}-\frac{1}{2}\mu
\end{equation*}%
where $\varepsilon +3p=-\mu $ and $\lambda _{1}+\lambda _{2}=\mu $ and these
functions are
\begin{equation*}
\lambda _{1}=-\int \frac{k_{0}^{2}}{k^{2}}b\frac{d^{4}k}{\left( 2\pi \right)
^{4}},\ \ \lambda _{2}=-\int \frac{\left\vert \mathbf{k}\right\vert ^{2}}{%
k^{2}}b\frac{d^{4}k}{\left( 2\pi \right) ^{4}}.
\end{equation*}%
By means of the use of the spherical coordinates $k_{0}^{2}/k^{2}=\cos
^{2}\theta $, $\left\vert \mathbf{k}\right\vert ^{2}/k^{2}=\sin ^{2}\theta $%
, and $d^{4}k=4\pi \sin ^{2}\theta k^{3}dkd\theta $ we get
\begin{equation*}
\lambda _{i}=\frac{n\left( a,r_{0}\right) }{4\beta _{i}}\left( 1-2\beta
_{i}\left( \frac{a}{r_{0}}\right) ^{2}f_{i}\left( \frac{a}{r_{0}}\right)
\right) ,
\end{equation*}%
where $\beta _{1}=1$, $\beta _{2}=1/3$, and $f_{i}$ is given by
\begin{equation*}
f_{\binom{1}{2}}\left( y\right) =\frac{2}{\pi }\int_{-1}^{1}\int_{0}^{\infty
}\sin \left( x\sin \theta \right) \frac{J_{1}\left( yx\right) }{yx/2}\binom{%
\cos ^{2}\theta }{\sin ^{2}\theta }dxd(\cos \theta ).
\end{equation*}%
For $\frac{a}{r_{0}}\ll 1$ we find
\begin{equation*}
f_{1,2}\left( \frac{a}{r_{0}}\right) \approx \left( 1+o_{1,2}\left( \frac{a}{%
r_{0}}\right) \right) \text{.}
\end{equation*}%
Thus, finally we find
\begin{equation*}
\varepsilon \simeq -\frac{n}{4}=-\frac{2\widetilde{n}}{3\pi a},\ \ \ p\simeq
\varepsilon \left( 1-\frac{4}{3}\left( \frac{a}{r_{0}}\right) ^{2}\right) .
\end{equation*}%
Which upon the continuation to the Minkowsky space gives the equation of
state in the form \footnote{%
An arbitrary gas of wormholes splits in fractions with a fixed $a$ and $%
r_{0} $.}
\begin{equation*}
p=-\left( 1-\frac{4}{3}\left( \frac{a}{r_{0}}\right) ^{2}\right) \varepsilon
,
\end{equation*}%
which in the case when $\frac{a}{r_{0}}\ll 1$ behaves like a cosmological
constant. However when $a\gg \ell _{pl}$ such a constant is extremely small
and can be neglected, while the leading contribution comes from the mean
curvature.

\subsection{Mean curvature}

In this subsection we consider the Minkowsky space. Then the simplest actual
wormhole can be described by the metric analogous to (\ref{wmetr}), e.g.,
see \cite{KS10}
\begin{equation}
ds^{2}=c^{2}dt^{2}-h^{2}\left( r\right) \delta _{\alpha \beta }dx^{\alpha
}dx^{\beta },  \label{wmetr3}
\end{equation}%
where $h\left( r\right) =1+\theta \left( a-r\right) \left( \frac{a^{2}}{r^{2}%
}-1\right) $. To avoid problems with the Bianchi identity and the
conservation of energy the step function should be also smoothed as in (\ref%
{wmetr}). The stress energy tensor which produces such a wormhole can be
found from the Einstein equation $8\pi GT_{\alpha }^{\beta }=R_{\alpha
}^{\beta }-\frac{1}{2}\delta _{\alpha }^{\beta }R$. Both regions $r>a$ and $%
r<a$ represent portions of the ordinary flat Minkowsky space and therefore
the curvature is $R_{i}^{k}\equiv 0$. However on the boundary $r=a$ it has
the singularity. Since the metric (\ref{wmetr3}) does not depend on time we
find
\begin{equation}
R_{0}^{0}=R_{\alpha }^{0}=0,\ \ \ R_{\alpha }^{\beta }=\frac{2}{a}\delta
\left( a-r\right) \{n_{\alpha }n^{\beta }+\delta _{\alpha }^{\beta
}\}+\lambda _{\alpha }^{\beta }
\end{equation}%
where $n^{\alpha }=n_{\alpha }=x^{\alpha }/r$ is the outer normal
to the throat $S^{2}$, and $\lambda _{\alpha }^{\beta }$ are
additional terms (e.g., see (\ref{Ricci}) and (\ref{R})) which in
the leading order are negligible upon averaging over some portion
of space $\Delta V\gtrsim a^{3}$. In the case of a set of
wormholes this gives in the leading order
\begin{equation}
R_{0}^{0}=R_{\alpha }^{0}=0,\ \ \ R_{\alpha }^{\beta }=\sum \frac{2}{a_{i}}%
\delta \left( a_{i}-\left\vert r-r_{i}\right\vert \right) \{n_{i\alpha
}n_{i}^{\beta }+\delta _{\alpha }^{\beta }\}
\end{equation}%
where $a_{i}$ is the radius of a throat and $r_{i}$ is the position of the
center of the throat in space and $n_{i}^{\alpha }=(x^{\alpha
}-r_{i}^{\alpha })/\left\vert r-r_{i}\right\vert $. In the case of a
homogeneous and isotropic distribution of such throats we find $R_{\alpha
}^{\beta }=\frac{1}{3}R\delta _{\alpha }^{\beta }$ (averaging over spatial
directions gives $\left\langle n_{\alpha }n^{\beta }\right\rangle =\frac{1}{3%
}\delta _{\alpha }^{\beta }$) where
\begin{equation}
R=-8\pi GT=\sum \frac{8}{a_{i}}\delta \left( a_{i}-\left\vert
r-R_{i}\right\vert \right) =32\pi \int a\widetilde{n}(a)da  \label{de}
\end{equation}%
where $T$ stands for the trace of the stress energy momentum tensor which
one has to add to the Einstein equations to support such a wormhole. It is
clear that such a source violates the weak energy condition and, therefore,
it reproduces the form of dark energy (i.e., $T=\varepsilon +3p<0$). If the
density of such sources (and respectively the density of wormholes) is
sufficiently high, then this results in the observed \cite{Ac} acceleration
of the scale factor for the Friedmann space as $\sim t^{\alpha }$ with $%
\alpha =\frac{2\varepsilon }{3(\varepsilon +p)}=\frac{2\varepsilon }{%
2\varepsilon +(\varepsilon +3p)}>1$, e.g., see also \cite{inf}. In terms of
the 4-dimensional density of wormholes $n=\frac{8}{3\pi a}\widetilde{n}$ we
get $R\sim a^{2}n\gg 8\pi Gn$ as $a\gg \ell _{pl}$ and, therefore, the
leading contribution indeed comes from the mean curvature.

\section{Estimates and concluding remarks}

Now consider the simplest estimates. Actual wormholes seem to be responsible
for the dark matter \cite{KS10,KS07}. Therefore, to get the estimate to the
number density of wormholes is rather straightforward. First wormholes
appear at scales when dark matter effects start to display themselves, i.e.,
at scales of the order $L\sim (1\div 5)Kpc$, which gives in that range the
number density
\begin{equation}
\widetilde{n}\sim \frac{1}{L^{3}}\sim (3\div 0.024)\times 10^{-65}cm^{-3}.
\label{n}
\end{equation}%
The characteristic size of throats can be estimated from (\ref{de}) $%
\varepsilon _{DE}$ $\sim $ $(G)^{-1}\widetilde{n}\bar{a}$. Since the density
of dark energy is $\varepsilon _{DE}/\varepsilon _{0}=\Omega _{DE}\sim $ $%
0.75$, where $\varepsilon _{0}$ is the critical density, then we find the
estimate
\begin{equation*}
\bar{a}\sim \frac{2}{3}(1\div 125)\times 10^{-3}R_{\odot }\Omega
_{DE}h_{75}^{2},
\end{equation*}%
where $R_{\odot }$ is the Solar radius, $h_{75}=H/(75km/(secMpc))$, and $H$
is the Hubble constant. We also recall that the background density of
baryons $\varepsilon _{b}$ generates a non-vanishing wormhole rest mass $%
M_{w}=\frac{4}{3}\pi \bar{a}^{3}R^{3}\varepsilon _{b}$ (where $R(t)$ is the
scale factor of the Universe and therefore $M_{w}$ remains constant) e.g.,
see \cite{KS10}. It produces the dark matter density related to the
wormholes as $\varepsilon _{DM}\simeq M_{w}\widetilde{n}$. The typical mass
of a wormhole $M_{w}$ is estimated as
\begin{equation*}
M_{w}\sim 1,7\times (1\div 125)\times 10^{2}M_{\odot }\Omega _{DM}h_{75}^{2},
\end{equation*}
where $M_{\odot }$ is the Solar mass. We point out that this mass has not
the direct relation to the parameters of the gas of wormholes. However it
defines the moment when wormhole throats separated from the cosmological
expansion. The above estimate shows that if wormholes form due to the
development of perturbations in the exotic matter, then this process should
start much earlier than the formation of galaxies.

Thus, we see that virtual wormholes should indeed lead to the
regularization of all divergencies in QFT which agrees with recent
results \cite{LQG}. Therefore, they form the local finite value of
the cosmological constant. In the absence of external classical
fields such a value should be exactly zero at macroscopic scales.
A some non-vanishing value for the cosmological constant appears
as the result of vacuum polarization effects in external fields.
Indeed, external fields form an additional distribution of virtual
wormholes which possess an exotic stress energy tensor (some kind
of dark energy). Only some part of it forms the cosmological
constant, while the rest reflects the symmetry of external fields
and possesses inhomogeneities. We assume that during the evolution
of our Universe inhomogeneities in the exotic matter develop and
may form actual wormholes. Although this problem requires the
further and more deep investigation we refer to \cite{KS12} where
the formation of a simplest wormhole-like object has been
considered. In other words, such polarization energy is reserved
now in a gas of actual wormholes. We estimated parameters of such
a gas and believe that such a gas may indeed be responsible for
both, dark matter and dark energy phenomena.



\end{document}